\documentclass[aps,epsfig,floats,showpacs,prb,preprint]{revtex4}

\usepackage[latin1]{inputenc} 
\usepackage{amsmath,amsthm}
\usepackage{amssymb} 
\usepackage{graphics}
\usepackage[dvips]{graphicx} 
\usepackage{epsfig} 
\usepackage{color}
\usepackage{epic} 
\usepackage{enumerate}

\newcommand\dps {\displaystyle }

\newcommand{\cD}{{\mathcal{D}}}
 
\newcommand{\cU}{{\mathcal{U}}}
\newcommand{\cP}{{\mathcal{P}}} 
\newcommand{\cS}{{\mathcal{S}}} 
\newcommand{\cH}{{\mathcal{H}}}
\newcommand{\rme}{{\mathrm{e}}} 
\newcommand{\OO}{{\mathrm{O}}}

\begin{document}

\preprint{APS/}

\title{Constant entropy sampling and release waves of shock
  compressions}

\author{Jean-Bernard Maillet,\footnote{corresponding author:
    jean-bernard.maillet@cea.fr} Emeric Bourasseau, Laurent Soulard,
  Jean Clérouin} \affiliation{CEA, DAM, DIF, F-91297 Arpajon, France}

\author{Gabriel Stoltz} \affiliation{ Universit\'e Paris Est, CERMICS,
  Project-team MICMAC, INRIA-Ecole des Ponts, 6 \& 8 Av. Pascal, 77455
  Marne-la-Vall\'ee Cedex 2, France }

\date{\today}

\begin{abstract}
We present or recall several equilibrium methods that allow to compute isentropic
processes, either during the compression or the release of the
material.  These methods are applied to compute the isentropic release
of a shocked monoatomic liquid at high pressure and
temperature. Moreover, equilibrium results of isentropic release are
compared to the direct nonequilibrium simulation of the same
process. We show that due to the viscosity of the liquid but also to
nonequilibrium effects, the release of the system is not strictly
isentropic.\\
\end{abstract}


\maketitle

\section{Introduction}

The exploration of the thermodynamic behavior of materials under
extreme conditions usually follows two paths corresponding to two
experimental devices: isothermal compressions and shock
compressions. Isothermal compressions are performed with diamond anvil
cell (DAC) techniques, and are used to compress materials up to very
high pressures, although with limited temperatures. On the other hand,
shock compression experiments investigate the high-pressure/high
temperature regions through the propagation of dynamic shock waves in
the system. Nevertheless, the thermodynamic domain available using
shock experiments remains limited to the so-called Hugoniot curve,
which is, by definition, the collection of thermodynamic states which
can be reached from a system at fixed initial conditions, with shocks
of increasing strengths. Such a constraint arises from the fact that
the system is assumed to satisfy Euler's fluid equations (\textit{i.e} inviscid 
Navier-Stokes equations),
and physically meaningful shocks therefore have to fulfill the
Rankine-Hugoniot conditions, which relate the thermodynamic parameters
of the fluid at rest, and the thermodynamic parameters of the shocked
material. Another constraint is that shock waves are adiabatic,
therefore leading to very large temperature increases in the material,
which limits its compressibility. The equations of state (EOS) used to
predict the material's behavior at the extreme conditions encountered
are often simple extrapolations of EOS fitted on available data, {\it
  i.e.} shock data and DAC data. It then appeared interesting to
enlarge the experimental domain of investigation of materials behavior
using dynamic compression set-ups, and particularly isentropic
compressions. Several experimental set-ups allow to load a pressure
ramp in a material. The first one is the high pulsed power (of which
the sandia Z machine and the High Explosive Pulsed
Power\cite{tasker05a, tasker05b} at LANL are good examples). The
second one consists in using an impactor with a varying density along
one direction, as proposed initially at the AIP - SWCM
conference.\cite{lyzenga81} A successful technique is to stack slices
of different materials, leading to the so-called PILLOW impactors at
Sandia,\cite{barker84} MIVAR impactors in France,\cite{perez88} and
more recently the FGM (Functionaly Graded Materials) impactors at
LLNL,\cite{nguyen06} allowing a real design of a
thermodynamic path as a succession of shock and release waves. The
last one concerns experiments of Barnes' type where the compression is
the consequence of the isentropic release of another material, as for
example detonation products.\cite{barnes74, tsypkin75}

Experiments involving isentropic compression are of great interest to
reach high compression states, or in geophysic applications to reach
states representative of the earth's core. Experiments involving a
precise evaluation of release waves in materials need also general
numerical methods to compute the states reached by isentropic
pocesses.

Up to now, simulations involving shock processes are rather well
developed due to the simplicity of the Hugoniot equations.  Those
studies are performed within the framework of statistical physics, see
Refs.~\onlinecite{frenkel,allen} for reference textbooks on molecular
simulation, and Ref.~\onlinecite{holian98} for reference works on
nonequilibrium simulation of shock waves.  Any state lying on the
Hugoniot can be reached from the reference state by searching for a
given compression the temperature for which the pressure and the total
energy of the system satisfies the Hugoniot relation.  The search can
be implemented in very efficient
manners.\cite{MMSRLGH00,MS08,Bourasseau07}  Those methods are now
adapted to classical molecular dynamics and Monte Carlo, as well as
quantum molecular dynamics.\cite{bernard02}

Such an easy method does not exist for isentropic processes. In this
paper, we present or recall several equilibrium methods which allow to follow
isentropic paths, both for classical or quantum molecular dynamics
simulations. We contrast these methods in terms of their precisions,
rigor and computational requirements. 
We compare the results obtained from {\it equilibrium} simulations 
with release
waves observed in {\it nonequilibrium} molecular dynamics. 
The comparison between equilibrium and nonequilibrium
methods therefore measures how isentropic the expansion of the
system is. 
It is expected that release waves of a perfect
non-viscous fluid are isentropic. For simple monoatomic fluids such as argon, 
it is often assumed that the release is isentropic, and
viscosity effects are neglected. Our results show that even in this simple case,
the release is not strictly isentropic and some corrections have to be taken into
account.
As a by-product of our study, we also explore more precisely the relationship
between the Hugoniot and the isentrope curves, from a numercial
viewpoint, but also giving a statistical physics proof of the
coincidence of the curves for small compressions (see Appendix~B).

\subsection*{Organization of the paper}

The paper is organized as follows. In
Section~\ref{sec:non_eq_methods}, we present the nonequilibrium method
used to simulate rarefaction waves, while some equilibrium methods for
constant entropy sampling are recalled in
Section~\ref{sec:eq_methods} - details of the practical implementaion
of the thermodynamic integration and precisions on the isentropic
integration are given in Appendix~A and C respectively). In Section~\ref{sec:numerical}, a
comparison of numerical results obtained in the case of release waves
in argon is performed, and we discuss whether release waves are
isentropic.


\section{Nonequilibrium simulations}
\label{sec:non_eq_methods}

\subsection{Microscopic description of physical systems}

We first recall how a system is described at the microscopic level
using statistical mechanics.\cite{Balian} Statistical physics is a
theory which allows to compute thermodynamic (macroscopic) properties
of a system knowing the microscopic interactions between its
constitutive elements.

Consider a microscopic system composed of $N$ particles, confined in a
simulation box $\mathcal{D} = [0,L_x] \times [0,L_y] \times [0,L_z]$.
The volume of the domain is denoted by $V = |\mathcal{D}| =
L_xL_yL_z$.  The system is characterized by the positions
$q=(q_1,\dots,q_N)$ and momenta $p=(p_1,\dots,p_N)$ of the particles,
which have masses $m_i$, and interact through a potential energy
function $U$.  The phase-space $\Omega$ is the collection of all
possible microscopic configurations $(q,p)$ of the system.

The central quantity describing the system is the Hamiltonian
\begin{equation}
  \label{eq:hamiltonian}
  H(q,p) = \sum_{i=1}^N \frac{p_i^2}{2m_i} + U(q_1,\dots,q_N),
\end{equation}
which gives the energy of a given microscopic configuration $(q,p)$.
Average thermodynamic properties of the system can be computed as
averages of functions of the microscopic variables $O(q,p)$ (the
so-called observables) with respect to the canonical measure at a
temperature $T$, for a given simulation box:
\begin{equation}
  \label{eq:canonical_average}
  \mathcal{O} = \langle O \rangle_{V,T} = \int_\Omega O(q,p) \,
  \pi_{V,T}(q,p) \, dq \, dp.
\end{equation}
The canonical measure associated with the
Hamiltonian~(\ref{eq:hamiltonian}) weights microscopic states
according to their energies using a Boltzmann weight:\cite{Balian}
\begin{equation}
  \label{eq:canonical_measure}
  \pi_{V,T}(q,p) = \frac{1}{Z_{V,T}} \textrm{e}^{-\beta H(q,p)},
  \qquad \beta^{-1} = k_{\rm B} T,
\end{equation}
where $k_{\rm B}$ is the Boltzmann constant, and the partition
function $Z_{V,T}$ is a normalization factor so
that~(\ref{eq:canonical_measure}) is indeed a probability measure:
\begin{equation}
  \label{eq:partition_fct}
  Z_{V,T} = \int_{\Omega} \textrm{e}^{-\beta H(q,p)} \, dq \, dp.
\end{equation}
We have indicated explicitly the dependence of the canonical
measure~(\ref{eq:canonical_measure}) and the partition
function~(\ref{eq:partition_fct}) on the temperature $T$ and the
volume $V$ since these will be the parameters allowed to vary in the sequel.

\subsection{Nonequilibrium simulation of release waves}
\label{sec:noneq_simulation}

Similarly to what has been proposed for the simulation of shock
waves,\cite{holian98} isentropic compressions or releases can be
simulated directly using Non-Equilibrium Molecular Dynamics (NEMD). A
straightforward numerical set-up to this end is simply to throw a low
speed piston towards the sample (creating a weak shock), and then
accelerating the piston in time.  Except this external forcing, the
system evolves according to the standard hamiltonian dynamics
\begin{equation}
  \label{eq:hamiltonian_dynamics}
  \left \{
  \begin{array}{ccl}
    \dot{q}_i & = & \dps \frac{p_i}{m_i}, \\ [10pt] 
    \dot{p}_i & = & \dps -\nabla_{q_i} U(q),
  \end{array}
  \right.
\end{equation}
which is integrated in time with the Verlet scheme.\cite{Verlet67} A
linear compression ramp would be obtained in the case where the
acceleration is constant in time.

To obtain release waves, a shock wave can be loaded in a sample;
when this shock wave is reflected when interacting with a free
surface, it transforms into a release wave, supposedly isentropic. In this study,
we start the release from an equilibrated state obtained from a
preliminary canonical simulation, using three dimensional periodic
boundary conditions. When the system is equilibrated, the periodic
boundary conditions are removed in the $x$ direction. Two release
waves are then created at the two free surfaces, and they propagate in
opposite directions towards the center of the box. This process is
illustrated in Figure~\ref{NEMDprocess}.

From the simulation data presented in Figure~\ref{NEMDprocess},
profiles of thermodynamic quantities 
(average densities, (kinetic) temperatures and
pressures) can be extracted and averaged
over thin slices. Moreover, the two release waves being symmetric,
their related profiles can be averaged. 
A superposition of the profiles, taken
at different times but projected back in the same thermodynamic diagram, is
then obtained and averaged over, leading to a single profile.

The accuracy of the computation increases with the system size: an
increase in the size of the transverse directions decreases the
uncertainties on the slice averages (thanks to a thermodynamic limit),
and an increase in the longitudinal direction allows to accumulate
more profiles in time, therefore reducing statistical errors.


\section{Equilibrium methods for constant entropy sampling}
\label{sec:eq_methods}

We present in this section three methods to compute the collection of
all states (in terms of their temperature, volume and pressure) which
have the same entropy as some reference state. These methods therefore
allow to draw a curve in the $(V,T)$ diagram (or in the $(P,T)$ or
$(P,V)$ diagrams), called the isentrope, and will be used as benchmark
methods in Section~\ref{sec:numerical} to check whether release waves
computed by NEMD simulations are indeed isentropic or not.  We
emphasize that, altough presented for the computation of isentropic
releases, all the methods described in this section may also be used
to determine isentropic compressions. 
We also recall a fourth method, used to obtain
the entropy of a system once the entropy of some reference state 
(such as the perfect gas) is fixed. 
The computational cost
of the latter method as well as its low accuracy for dense states prevented
us from applying it to enough points to obtain an entire isentrope
curve, and we therefore limited its use to a consistency check on the 
results obtained with the other methods.

\subsection{Thermodynamic integration}
\label{sec:TI}

The entropy of the system varies when the simulation conditions are
changed. Here, we consider that the states visited by the release wave
are a succession of local equilibrium states, which can be described
within the canonical ensemble as given by statistical
physics. Therefore, the state of the system is defined by two
parameters, its volume (equivalently, the density) and its
temperature.

\subsubsection{Variables indexing the variations.}
Consider a general transformation in which both the volume accessible
to the system (equivalently, the density) and the temperature are
varied. We restrict ourselves to variations of the domain in one
spatial direction only, to model the anisotropic behavior of release
waves. Assuming that the state of the system at rest can be described
by some cubic simulation box with periodic boundary conditions, the
volume under compression may be indexed by a variable $\lambda_1$, so
that the associated simulation domain $\mathcal{D}(\lambda_1) =
[0,(1+\lambda_1)L_x] \times [0,L]^2$ has a volume
\[
V(\lambda_1) = (1+\lambda_1) L_x L^2.
\]
Notice that we consider $L_x \not = L$
since we may start from a uniaxially compressed state. The
temperature variations are indexed by a parameter $\lambda_2$:
\[
T(\lambda_2) = (1 + \lambda_2 \delta T) T,
\] 
for some reference temperature $T$ and a given relative temperature
variation $\delta T$, the temperature variation being therefore 
$\Delta T = T \delta T$. The reference inverse temperature is still
$\beta^{-1} = k_{\rm B}T$.  The particular case where only the
temperature is changed (while the volume is kept constant) corresponds
to $\lambda_1$ constant, while isothermal transformations are characterized
by $\lambda_2$ remaining constant. Expansions correspond to $\lambda_1 > 0$.

\subsubsection{Parametrization of the isentrope curve.}
The isentrope is the locus of the points in the
$(\lambda_1,\lambda_2)$ space such that the entropy normalized by the
Boltzmann factor
\begin{equation}
  \label{eq:entropy_constant}
  \frac{\mathcal{S}}{k_{\rm B}} = \frac{\mathcal{U} -
    \mathcal{F}}{k_{\rm B}T}
\end{equation}
is constant, $\mathcal{F}$ denoting the free energy of the system, and
$\mathcal{U}$ its energy.  This thermodynamic relation can be
converted into an equivalent formula in the framework of statistical
physics, which is much more convenient from a computational viewpoint:
\[
\mathcal{U} \equiv \mathcal{U}(T,V) = \langle H \rangle_{V,T},
\]
where the canonical average is defined in
Eq.~\eqref{eq:canonical_average}, and 
\[
\mathcal{F} \equiv \mathcal{F}(T,V) = - k_{\rm B} T \ln \int_\Omega
        {\rm e}^{-\beta H(q,p)} \, dq \, dp.
\]
We start from some reference state described by the parameters
$(\lambda_1,\lambda_2)=(0,0)$. The statistical
physics reformulation of the requirement that 
\eqref{eq:entropy_constant} be constant
is then
\begin{eqnarray*}
  \cS(\lambda_1,\lambda_2) - \cS(0,0) & = & \frac{1}{k_{\rm B}T(\lambda_2)} \langle H
  \rangle_{V(\lambda_1),T(\lambda_2)} - \frac{1}{k_{\rm B}T(0)} \langle H \rangle_{V(0),T(0)} \\ 
  & & + \ln \left ( \frac{\dps
    \int_{\Omega(\lambda_1)} \rme^{-H(q,p)/k_{\rm B}T(\lambda_2)} \, dq
    \, dp} {\dps \int_{\Omega(0)}
    \rme^{-H(q,p)/k_{\rm B}T(0)} \, dq \, dp} \right)=0.
\end{eqnarray*}
In this expression, the phase-space $\Omega(\lambda_1)$ is the
collection of all possible microscopic configurations of the system
associated with a domain $\cD(\lambda_1)$ of volume $V(\lambda_1)$.

\subsubsection{Numerical implementation}

To determine the isentrope curve, we compute the entropy variation
along a given path in $(\lambda_1,\lambda_2)$ space going through the reference initial state,
and search for the point such that the entropy difference
with this state is 0. A simple choice is illustrated in
Fig.~\ref{fig:TI_path}. It consists in performing
\begin{enumerate}[(i)]
\item an isothermal rarefaction, going from the initial compressed
  state $(0,0)$ to an
  intermediate state $(\lambda_1,0)$ with
  $\lambda_1 \geq 0$;
\item in a second step, an isochore cooling, going from the
  intermediate state $(\lambda_1,0)$ to some final
  state $(\lambda_1,\lambda_2)$, resorting to a maximal temperature
  difference $\lambda_2 \Delta T < 0$ large
  enough.
\end{enumerate}
The idea is that, in general, the first part of the transformation
increases the entropy of the system (since more space becomes
available for the particles), while the entropy decreases in the
second part (since the temperature decreases).  Of course, more
general paths, with joint variations of $\lambda_1$ and $\lambda_2$,
could be considered.

The energies $\langle H \rangle_{V(\lambda_1),T(\lambda_2)}$ are computed using
standard sampling strategies, while the remainder in the expression of 
$\cS(\lambda_1,\lambda_2) - \cS(0,0)$,
a ratio of partition functions, is estimated using standard techniques 
for free-energy calculations. This is detailed in Appendix~A.

We emphasize that this procedure is time consuming since it requires
many equilibrium samplings to obtain one point on the curve. It is
however exact (up to statistical errors and discretization errors in
the integrals defining $A$), and can be straightforwardly parallelized
since the equilibrium samplings required are independent.

\subsection{Successive hugoniostat simulations}

The variations of macroscopic quantities across a shock interface are
governed by the Rankine-Hugoniot relations, which relate the jumps of
the quantities under investigation (pressure, density, velocities) to
the velocity of the shock front.  The third Rankine-Hugoniot
conservation law for the Euler equation governing the hydrodynamic
evolution of the fluid reads (macroscopic quantities are denoted by
curly letters)
\begin{equation}
  \label{eq:RH3}
  \mathcal{H} = \mathcal{U} - \mathcal{U}_0 - \frac12 (\mathcal{P} + \mathcal{P}_0)
  (V_0-V) = 0.
\end{equation}
In this expression, $\mathcal{U}$ is the internal energy of the fluid,
$\mathcal{P}$ its pressure, and $V$ its volume. The
subscript $0$ refers to the initial state (the pole), the other quantities
are evaluated at a state obtained from some shock compression, after
equilibration.  The Hugoniot curve corresponds to all the possible
states satisfying~(\ref{eq:RH3}). In practice, the collection of these
states may be computed by nonequilibrium simulations with shocks of
different strengths, inducing various compressions.

Alternatively, small equilibrium simulations may be used, relying on
the statistical physics reformulation of the Hugoniot relation:
\begin{eqnarray}
  \label{eq:Hugoniot_statistical_physics}
  \cH(\lambda_1,\lambda_2) - \cH(0,0) & = & \langle H \rangle_{V(\lambda_1),T(\lambda_2)} -
  \langle H \rangle_{V(0),T(0)}
  \nonumber \\ 
  & & \hspace{-2cm} + \frac{\lambda_1}{2} V(0) \left ( \langle P_{xx}
  \rangle_{V(\lambda_1),T(\lambda_2)} + \langle P_{xx}
  \rangle_{V(0),T(0)} \right ) = 0.
\end{eqnarray}
The $xx$ component of the pressure tensor is, for a simulation domain
of volume $V(\lambda_1)$,
\begin{equation}
  \label{eq:xx_pressure}
  P_{xx}(q,p) = \frac{1}{V(\lambda_1)} \sum_{i=1}^N
  \frac{p_{i,x}^2}{m_i} - q_{i,x} \, \partial_{q_{i,x}} U(q).
\end{equation}
For a given variation of the volume for instance (indexed by
$\lambda_1$), the variation $\lambda_2 T \delta T$ of the temperature
is sought for, using for instance the techniques described in
Refs.~\onlinecite{MMSRLGH00,MS08}.

The Hugoniot curve does not have {\it a priori} any relationship with
the isentrope curve. However, it can be shown that the entropy
variation along the Hugoniot curve is negligible up to terms of order
three in the volume variable; the Hugoniot and the isentropic curves
are osculatory. We present in Appendix~B two proofs, the standard proof
based on thermodynamic relations, and a new proof fully relying on a
statistical physics reformulation.

The good agreement between the Hugoniot and the isentrope for small
compressions and/or expansions can be used to compute the isentropic
curve as a succession of weak shocks or weak releases, this
approximation getting more accurate as the shock compressions are
weakened. The only parameter left in
this method is the relative volume change $\delta V/V =
\lambda_1^{n+1}-\lambda_1^n$ during the instantaneous compressions or
releases. We used the Hugoniostat
method\cite{MMSRLGH00,MS08} to compute a sequence of states $(\lambda_1^n,\lambda_2^n)$
such that $\mathcal{H}(\lambda_1^{n+1},\lambda_2^{n+1}) = \mathcal{H}(\lambda_1^n,\lambda_2^n)$,
the corresponding thermodynamic properties at these states being obtained as a by-product
of the simulation.

\subsection{Isentropic integration}

Another way to perform thermodynamic integration along an isentropic
path has been proposed by Desjarlais.\cite{desjarlais06} The method
relies on the equilibrium evaluation of $\dps \frac{\partial
  \cP}{\partial \cU}$ (see Eq.~\eqref{eq:entropy_integration} below).
It could be
applied to a system where the pressure is not isotropic upon 
replacing the pressure observable 
by the $xx$ component of the pressure tensor.

The total differential of the entropy can be written as:
\begin{equation}
  \label{eq:Sdifferential} 
  d\cS = \frac{\partial \cS}{\partial T} \bigg |_V dT +
  \frac{\partial \cS}{\partial V} \bigg |_T dV.
\end{equation}
For constant entropy processes,
\[
\frac{\partial \cS}{\partial T} \bigg |_V = \frac{1}{T}
\frac{\partial \cU}{\partial T} \bigg |_V
= -\frac{\partial \cP}{\partial T} \bigg |_V, 
\qquad 
\frac{\partial \cS}{\partial V} \bigg |_T = \frac{\partial \cP}{\partial T} 
\bigg|_V,
\]
so that, along the isentrope,
\[
\frac{dT}{T} = -\frac{\dps \frac{\partial \cP}{\partial T} \bigg
  |_V} {\dps \frac{\partial \cU}{\partial T} \bigg |_V} \, dV =
-\frac{\partial \cP}{\partial \cU}\bigg |_V dV.
\]
This equation can be integrated as
\begin{equation}
  \label{eq:entropy_integration} 
  \frac{T_2}{T_1}=\exp \left (- \int_{V_1}^{V_2}{\frac{\partial
      \cP}{\partial \cU} \bigg |_V dV} \right),
\end{equation}
giving the temperature $T_2$ at which the system at volume $V_2$
has the same entropy as the system in the reference state
$(T_1,\cP_1)$.
This formula is evaluated in practice by discretizing the integral appearing
in the exponential, and approximating the integrand using standard 
canonical sampling procedues. We refer to Appendix~C for more precisions.

\subsection{Evaluation of the entropy based on the chemical potential}

This technique, which can be used only for systems in a fluid phase,
follows the classical methodology of computing the free
energy $\mathcal{F}$ of a system starting from the thermodynamic
relation:\cite{kofke97}
\begin{equation}
\mathcal{F} = \cU - T\cS = N \mu - \cP V,
\end{equation}
where $\mu$ is the chemical potential, defined in the canonical ensemble as
\begin{equation}
  \label{eq:chemical_pot}
  \mu = \frac{\partial \mathcal{F}}{\partial N}.
\end{equation} 
In the case of a canonical simulation, all thermodynamic quantities
are functions of the volume and the temperature, so that
\begin{equation}
\cS = \frac{\cU(T,V) + N \mu(T,V) - \cP(T,V) V}{T}.
\end{equation} 
This expression allows to compute the absolute entropy of the system
provided the chemical potential is known,\cite{lisal06} the average
pressure and energy being computed using standard sampling techniques.
The chemical potential is estimated using the Widom insertion method.


\section{Numerical results for release waves}
\label{sec:numerical}

We compare in this section the results for the different techniques
presented in Sections~\ref{sec:non_eq_methods}
and~\ref{sec:eq_methods}, for a release in a Lennard-Jones system
(argon). The aim is to assess whether the release is indeed
isentropic, and also to demonstrate that 
approximate equilibrium computations for small systems 
(successive Hugoniostat, isentropic integration) can
approximate the isentrope curve obtained from the more rigourous and
costly thermodynamic integration technique.

\subsection{Numerical parameters}

\subsubsection{Initial state.} 

We consider argon in an initial shocked state, 
located on the Hugoniot curve for a compression such that $L_x = c L$ with 
$c=0.65$, 
and corresponding to $T=1758$~K and
$P = 1.7 \times 10^{10}$~Pa. At these thermodynamic conditions, the system is in
a liquid state.  The interactions within noble gas atoms are
well-described by a Lennard-Jones potential:
\[
V(q_1,\dots,q_N) = \sum_{1 \leq i < j \leq N} v(|q_i-q_j|), \qquad
v(r) = 4\varepsilon \left ( \left(\frac{\sigma}{r}\right)^{12}
-\left(\frac{\sigma}{r}\right)^{6} \right ).
\]
In the case of Argon, $\varepsilon/k_{\rm B} = 120$~K and
$\sigma=3.405$~\AA.  The cut-off radius for the Lennard-Jones
interaction is here $r_{\rm cut} = 2.5 \sigma$.

\subsubsection{Nonequilibrium simulations.}
 
In order to reach this initial state before performing the NEMD release, 
a preliminary hugoniostat simulation is run
for a system of $50 \times 50 \times 500$ unit cells, using periodic
boundary conditions. Then, the boundary conditions in the longitudinal
direction are removed, and the system evolves according to the Hamiltonian
dynamics. Profiles of thermodynamic quantities are
computed every 0.25 ps for the post-processing procedure described 
at the end of Section~\ref{sec:noneq_simulation}.

\subsubsection{Equilibrium simulations}

Equilibrium computations have been performed with 
a system composed of $N=4000$ atoms, starting in a FCC crystal
geometry before melting, using periodic boundary conditions in all
directions.

\paragraph{Thermodynamic integration.}
As shown in Section~\ref{sec:TI}, 
the search of states having the same entropy can be performed using
thermodynamic integration, which amounts to performing many
equilibrium simulations.
The canonical sampling for a given set of
parameters $(\lambda_1,\lambda_2)$ is done with a Langevin
dynamics for $N_{\rm steps} = 2^{17}$ time steps, with $\Delta t = 2 \times
10^{-15}$~s, and a friction coefficient $\gamma = 10^{13}$~s$^{-1}$.

First, the entropy variation along the isothermal release is computed,
with canonical samplings along the path $(0,0) \to (0,\lambda_1)$ 
with $\lambda_1 = 0.54$ (using $M+1 = 15$ 
states). Then, for each compression of interest, the
isochore cooling is performed using temperature steps $\Delta T =
-25$~K for expansions $\lambda_1 \leq 0.25$, and $\Delta T =
-50$~K for states $\lambda_1 \geq 0.25$ (these paths can be restated
in terms of $\lambda_2 \in [0,1]$ upon considering a temperature
modification $\Delta T$ depending on the compression).
The numerical
integration for computing the value of $A$ is finally performed using the
trapezoidal rule.

Error estimates on the canonical samplings are obtained with block
averaging.\cite{FP89} In all the cases considered, the statistical
error (as measured using the 95\% confidence interval associated with
the variance computed from block-averaging) is inferior to
1\%. Therefore, the entropy difference is computed within 1\%
errors. For fixed $\lambda_1$, the state $\lambda_2$ such that
$\cS(\lambda_1,\lambda_2)$ is constant is then known with an error depending on
the local value of the partial derivative of $\cS$ with respect to
$\lambda_2$.  This error can immediately be reformulated as an error
on the estimated temperature. The error on the computed pressure is
the error arising from the error on the state $\lambda_2$, plus
the sampling error. It is found to be at most 2\%.

\paragraph{Successive Hugoniostat.}
Successive Hugoniostat simulations have been performed with a Langevin
version of the Hugoniostat method (see Eq.~(11) in
Ref.~\onlinecite{MSpreprint}, with the parameters $\xi =
10^{12}$~s$^{-1}$ and $\nu = 10^{12}$~s$^{-1}$).  Trajectories of
$N_{\rm steps} = 50,000$ timesteps at each compression are considered,
with a timestep $\Delta t = 5 \times 10^{-16}$~s. The relative volume
change $\delta V/V_0$ from one point on the curve to another is set to
0.01.

\paragraph{Isentropic integration.} See Appendix~C.

\paragraph{Entropy evaluation.}
The test particle insertion method used to evaluate the chemical
potential requires many more iterations than the other 
equilibrium techniques. In the same framework as for
isentropic integration (see Appendix~C), 
$N_{\rm steps} = 5 \times 10^8$ iterations were needed to
obtain a satisfactory convergence. The statistical error 
on the calculated entropy (as measured
using the 95\% confidence interval associated with the variance
computed from block-averaging) is estimated to be
inferior to 1.2 \%.

\subsection{Discussion of the numerical results}

Release waves are presented in Figures~\ref{P_rho}-\ref{T_rho} in three
different diagrams, $(P,\rho)$,$(P,T)$ and $(T,\rho)$.  It can clearly
be seen that the results coming from the three equilibrium techniques of
isentropic simulations are very close. This shows that provided the
relative volume change parameter is carefully chosen in either the
successive hugoniostat method or the isentropic integration, the
propagating error remains at a low level; these methods can then be as
accurate as the more rigorous and costly method of thermodynamic
integration. Moreover, evaluating the chemical potential, we have computed
absolute value of the entropy at three densities, $\rho=2780$
kg.m$^{-3}$, $\rho=2190$ kg.m$^{-3}$, and $\rho=1806$ kg.m$^{-3}$. The
corresponding values, $83.2 \pm 0.95$~J.mol$^{-1}$, $83.1 \pm 0.42$~J.mol$^{-1}$, 
$83.2 \pm 0.18$~J.mol$^{-1}$, confirm that the entropy is
indeed constant (within the error bars) on the calculated curve,
validating once again the different methods.

The comparison with the
results of the expansion of the liquid using nonequilibrium MD is
also fruitful. The overall agreement is fair enough, 
which means that release waves are indeed almost isentropic.
However, it can be noticed that the temperature is not predicted correctly.
While the different curves look very similar in a
($P,\rho$) diagrams, some discrepancies appear in the ($T,P$) diagram, 
which are even more obvious in the ($T,\rho$) diagram. 
Indeed, for the latter diagram, the observed
temperatures around the final density
are greater than the error bars. The
thermodynamic path followed by the system during its release exhibits
systematically a higher temperature than the one of an isentropic
process. This means that the release of a monoatomic 
liquid is not strictly isentropic, as is sometimes expected or assumed. 

Recall however that a release is expected to be isentropic only 
for non-turbulent flows of non-viscous
fluids. In the case considered here, the fluid has a finite, non zero viscosity,
and therefore dissipates energy under the form of heat. 
As a consequence, the temperature should be higher than for an isentropic
release. A tentative of evaluation of this effect is presented
below. On the Hugoniot curve,\footnote{Since the Hugoniot curve and the 
isentrope are close enough, we assume that the 
impact of the viscosity has roughly the same amplitude on the
isentrope curve.} viscous effects can be introduced in the
Hugoniot relation\cite{fickett} by means of the "viscous pressure"
$\pi$ as
\begin{equation}
  \frac{1}{2} \pi (\mathcal V_0 -\mathcal V)= \mathcal{U} -
  \mathcal{U}_0 - \frac12 (\mathcal{P} + \mathcal{P}_0)
(V_0-V),
\end{equation}
where $\pi$ is defined as
\begin{equation}
  \pi = -\nu \frac{du}{dx},
\end{equation}
$\nu$ being the fluid viscosity and $\dps \frac{du}{dx}$ the velocity
gradient. Taking the viscosity of the argon fluid at $T=700$~K and 
$P = 1$~GPa
(the most extremes conditions of available thermodynamic tables), and
considering an average velocity gradient (taken during the fluid
release), we find a temperature elevation of a few
Kelvins. Considering that the pressure is much higher in our simulation, and
therefore that the viscosity should be also greater, the actual
temperature increase due to the finite viscosity 
should rather be of the order of a few tens of Kelvins, 
which is consistent with what can be observed in our numerical results. 

Finally, a purely nonequilibrium effect has been
observed during the NEMD simulations, that also leads to a temperature
increase in the system. Indeed, gradients of thermodynamic and
kinematic quantities are large at the first stages of the release, when
the hot and dense material is in contact with void. The thermodynamic
path followed by the system at those early stages of the simulation does
not correspond to the thermodynamic path followed when the release has reached
its self-similarity regime. Some equilibration time is needed for some 
steady-state regime to be reached. We evaluated
this time to be around 5 picoseconds.


\section{Conclusion}

We have presented or recalled several equilibrium methods to compute
isentropic processes in the high pressure regime, either for
compressions or releases. These methods, although very different in
nature, lead to similar results when applied to the release of a
monoatomic liquid.

We have then compared release waves computed with these equilibrium methods
with the nonequilibrium simulation of the release process. The results
show that the release is almost, but not strictly isentropic, the
system's temperature being systematically greater than the one of
the isentropic process. This is the consequence of two effects.
First, the fluid actually has a finite viscosity and therefore 
dissipates heat, leading to a temperature increase. To our knowledge, this
is the first time that this effect has been quantified rigorously using
molecular dynamics simulations.
Moreover, the thermodynamic path
followed by the system during its release takes some time to reach a
converged profile. We anticipate that these effects will be enhanced
in the case of a more complex fluid, for example in the case of a the
release of detonation product. 
Therefore, the assumption that release waves are isentropic should
be carefully verified in each case.


\appendix

\section{Practical implementation of the thermodynamic integration}

\subsection{Reformulation of the problem in a fixed geometry}

From a computational viewpoint, it is more convenient to work with a
fixed simulation domain. For instance, the unperturbed domain $V(0)$
may be used to fix the geometry of the system. The volume variations
are then rephrased as variations in the interaction scale between the
particles in the direction of compression or release. In the same
vein, the temperature may be kept constant, upon rescaling the
interactions strengh by a factor depending on the temperature
variation.  Introducing the rescaled potential energy for a
configuration $q=(x,y,z)$:
\[
U_{\lambda_1,\lambda_2}(q) = \frac{1}{1+\lambda_2 \delta T}
U((1+\lambda_1)x,y,z).
\]
and the associated Hamiltonian
\[
H_{\lambda_1,\lambda_2}(q,p) = U_{\lambda_1}(q) + \frac{1}{1+\lambda_2
  \delta T} \sum_{i=1}^N \frac{p_i^2}{2m_i},
\]
canonical averages for a volume $V(\lambda_1)$ at a temperature
$T(\lambda_2)$ can be reformulated as canonical averages in terms of
the rescaled Hamiltonian $H_{\lambda_1,\lambda_2}$ at the reference
state at volume $V(0)$ and temperature $T(0)$. More precisely,
\[
\langle H \rangle_{V(\lambda_1),T(\lambda_2)} = \frac{3N}{2} k_{\rm
  B}T(\lambda_2) + (1+\lambda_2\delta T) \langle \langle
U_{\lambda_1,\lambda_2} \rangle \rangle_{\lambda_1,\lambda_2},
\]
where
\[
\langle \langle f \rangle \rangle_{\lambda_1,\lambda_2} = \frac{\dps
  \int_{\Omega(0)} f(q,p) \,
  \rme^{-H_{\lambda_1,\lambda_2}(q,p)/k_{\rm B}T(0)} \, dq \, dp}
        {\dps \int_{\Omega(0)}
          \rme^{-H_{\lambda_1,\lambda_2}(q,p)/k_{\rm B}T(0)} \, dq \,
          dp}.
\]
It is then easily seen that
\begin{eqnarray}
  \label{eq:expression_cS}
  \cS(\lambda_1,\lambda_2) - \cS(0,0) & = & \frac{3N}{2} 
  \ln(1 + \lambda_2 \delta T) + N \ln(1+\lambda_1) \nonumber \\ 
  & & + \beta \left [ \langle \langle U_{\lambda_1,\lambda_2} \rangle
    \rangle_{\lambda_1,\lambda_2} - \langle \langle U_{0,0} \rangle \rangle_{0,0}\right ] \\ & & +
  A(\lambda_1,\lambda_2),
  \nonumber
\end{eqnarray}
with
\[
A(\lambda_1,\lambda_2) = \ln
\left ( \frac{\dps \int_{V(0)^N} \rme^{-\beta
    U_{\lambda_1,\lambda_2}(q)} \, dq} {\dps \int_{V(0)^N}
  \rme^{-\beta U_{0,0}(q)} \,
  dq} \right ).
\]
In the above expression of the entropy difference, the first line is
the ideal gas contribution to the entropy difference. As a consistency
check, we can verify that the entropy increases when the volume or the
temperature is increased, as expected.  The terms on the second and
third lines in Eq.~\eqref{eq:expression_cS} are the ``excess''
contributions associated with the potential interaction energy.

\subsection{Numerical evaluation of the different terms}

To estimate $\cS$, two quantities are required:
\begin{enumerate}[(i)]
\item averages $\langle \langle \cdot \rangle
  \rangle_{\lambda_1,\lambda_2}$ with respect to the Hamiltonian
  $H_{\lambda_1,\lambda_2}$ are computed using standard sampling
  techniques such as a Langevin dynamics at an inverse temperature
  $\beta$, implemented using the so-called BBK algorithm.\cite{BBK84}
  Of course, many other sampling techniques could be used to estimate
  this canonical average, in particular Nos\'e-Hoover
  dynamics\cite{Nose84,Hoover85} or Metropolis-Hastings
  schemes\cite{MRRTT53,hastings70} (see Ref.~\onlinecite{CLS07} for a
  mathematical review on sampling methods in the context of molecular
  simulation);
\item the term $A(\lambda_1,\lambda_2)$ requires more care in its
  estimation. Since this term is a ratio of partition functions,
  standard techniques used for the computation of free energy
  differences may be used. We resorted to thermodynamic
  integration,\cite{Kirkwood35} in which case the function is
  rewritten as the integral of some canonical averages:
  \[
  A(\lambda_1,\lambda_2) =
  \int_{0}^{\lambda_1} \frac{\partial
    A}{\partial_{\lambda_1}}(x,0) \, dx +
  \int_{0}^{\lambda_2} \frac{\partial
    A}{\partial_{\lambda_2}}(\lambda_1,x) \, dx,
  \]
  with
  \begin{equation}
    \label{eq:derivatives_A_2}
    \frac{\partial A}{\partial \lambda_2}(\lambda_1,\lambda_2) =
    \beta \, \frac{\delta T}{1+\lambda_2\delta T} \, \langle \langle
    U_{\lambda_1,\lambda_2} \rangle \rangle_{\lambda_1,\lambda_2},
  \end{equation}
  and
  \begin{equation} 
    \label{eq:derivatives_A_1}
    \frac{\partial A}{\partial \lambda_1}(\lambda_1,\lambda_2) = 
    \big \langle \big \langle \, x \cdot \nabla_x
    U((1+\lambda_1)x,y,z) \, \big \rangle \big
    \rangle_{\lambda_1,\lambda_2}.  
  \end{equation}
\end{enumerate}
In conclusion, the numerical procedure consists in first estimating
the derivatives of the function $A$ and the average potential energy,
for as many points as required on the thermodynamic path chosen.
Approximations of $\cS$ can then be obtained thanks to
\eqref{eq:expression_cS}, after a numerical integration to obtain $A$.
The entropy difference along the path is then plotted, and fixing the
volume change $\lambda_1$, the temperature variation is chosen such
that the entropy difference is 0. This determines $\lambda_2$ as a
function of $\lambda_1$.

\section{Relationship between the Hugoniot and the isentrope curves at the pole}

We present in this Appendix two proofs of the fact that the isentrope
curve and the Hugoniot agree at order 3 in the volume change. The
first one is a standard thermodynamic proof, but the second one, based
on statistical physics relations, is new to the best of our knowledge.

\subsection{Standard thermodynamic proof}

For the sake of completeness, we reproduce here the proof of 
Ref.~\onlinecite{thouvenin}.  From the thermodynamic relation $Td\cS = d\cU + \cP \, dV$
and from the Hugoniot relation $\cU = \cU_0 + \frac{1}{2}
(\cP+\cP_0)(V_0-V)$, the entropy variation along the Hugoniot curve
can be computed. One derivation leads to
\begin{equation}
  T \left ( \frac{d\cS}{dV} \right)_{\rm Hug} = \cP+\left (
  \frac{d\cU}{dV} \right)_{\rm Hug}
  =\frac{1}{2}(\cP-\cP_0)+\frac{1}{2}(V_0-V)\left ( \frac{d\cP}{dV}
  \right)_{\rm Hug}.
\end{equation}
A second derivation gives
\begin{equation}
T\frac{d^2\cS_{\rm Hug}}{dV^2}+\frac{d\cS_{\rm Hug}}{dV} \frac{dT_{\rm
    Hug}}{dV}=\frac{1}{2}(V_0-V) \frac{d^2\cP_{\rm Hug}}{dV^2}.
\end{equation}
With a final derivation,
\begin{equation}
T\frac{d^3\cS_{\rm Hug}}{dV^3}+2\frac{d^2\cS_{\rm Hug}}{dV^2}
\frac{dT_{\rm Hug}}{dV}+\frac{d\cS_{\rm Hug}}{dV} \frac{d^2T_{\rm
    Hug}}{dV^2} =-\frac{1}{2}\frac{d^2\cP_{\rm
    Hug}}{dV^2}+\frac{1}{2}(V_0-V)\frac{d^3\cP_{\rm Hug}}{dV^3}.
\end{equation}
At the initial state (denoted with a subscript 0), that is, in the
limit $V \to V_0$, it holds:
\begin{eqnarray*}
\left (\frac{d\cS_{\rm Hug}}{dV} \right )_0 & = & 0, \\ 
\left(\frac{d^2\cS_{\rm Hug}}{dV^2} \right )_0 & = & 0, \\ 
T_0\left(\frac{d^3\cS_{\rm Hug}}{dV^3} \right )_0 & = & - \frac12
\left(\frac{d^2\cP}{dV^2} \right )_0 \not = 0.
\end{eqnarray*}
The entropy variation along the Hugoniot curve is therefore of order
three in volume, and the Hugoniot and the isentropic curves are
osculatory.

\subsection{A statistical physics derivation}

Without loss of
generality (and for notational simplicity), 
we may set $\cH(0,0) = \cS(0,0) = 0$ since we are only interested in 
differences of $\mathcal{F}$ and $\mathcal{S}$.

\subsubsection{Some useful relations.}
The derivatives of the function $A$ are useful for comparing the
Hugoniot and the isentrope relations.  The average $xx$ component of
the pressure tensor for the volume $V(\lambda_1)$ and the temperature
$T(\lambda_2)$ is obtained by averaging the observable
\[
P_{xx}(q,p) = \frac{1}{V(\lambda_1)} \left ( N k_{\rm B}T(\lambda_2) -
x \cdot \nabla_x U(q) \right ).
\]
Therefore,
\begin{eqnarray*}
  \langle P_{xx} \rangle_{V(\lambda_1),T(\lambda_2)} & = &
  \frac{N}{(1+\lambda_1) V(0)} \, k_{\rm B}T(\lambda_2) -
  \frac{1}{V(\lambda_1)} \frac{\dps \int_{V(\lambda_1)^N} x \cdot
    \nabla_x U(q) \, \rme^{-U(q)/k_{\rm B}T(\lambda_2)} \, dq} {\dps
    \int_{V(\lambda_1)^N} \rme^{-U(q)/k_{\rm B}T(\lambda_2)} \, dq} \\ 
  & = & \frac{N}{\beta V(0)} \, \frac{1+\lambda_2\delta T}{1+\lambda_1} -
  \frac{1+\lambda_1}{V(\lambda_1)} \frac{\dps \int_{V(0)^N} x \cdot
    \nabla_x U((1+\lambda_1)x,y,z) \, \rme^{-\beta
      U_{\lambda_1,\lambda_2}(q)} \, dq} {\dps \int_{V(0)^N}
    \rme^{-\beta U_{\lambda_1,\lambda_2}}}.
\end{eqnarray*}
This shows that, using~\eqref{eq:derivatives_A_1},
\[
\langle P_{xx} \rangle_{V(\lambda_1),T(\lambda_2)} = \frac{N}{\beta
  V(0)} \, \frac{1+\lambda_2\delta T}{1+\lambda_1} +
\frac{1+\lambda_2\delta T}{\beta V(0)} \frac{\partial
  A}{\partial_{\lambda_1}}(\lambda_1,\lambda_2).
\]

\subsubsection{Hugoniot curve.} 
With the above computations, it is easily seen that the Hugoniot
relation \eqref{eq:Hugoniot_statistical_physics} can be restated as
\begin{eqnarray*}
  \beta \cH(\lambda_1,\lambda_2) & = & \frac{3N}{2} \lambda_2 \delta T +
  \beta \left [ (1+\lambda_2\delta T) \langle \langle
    U_{\lambda_1,\lambda_2} \rangle \rangle_{\lambda_1,\lambda_2} -
    \langle \langle U_{0,0} \rangle \rangle_{0,0} \right ] \\ 
  & & +
  \frac{N\lambda_1}{2} \left ( \frac{1+\lambda_2\delta T}{1+\lambda_1} +
  1 \right ) + \frac{\lambda_1}{2} \left (\frac{\partial
    A}{\partial_{\lambda_1}}(0,0) + (1+\lambda_2\delta T) \frac{\partial
    A}{\partial_{\lambda_1}}(\lambda_1,\lambda_2)\right ).
\end{eqnarray*}

\subsubsection{Comparison between the Hugoniot and the isentrope.}
We now Taylor expand the difference $\beta \cH(\lambda_1,\lambda_2) -
\cS(\lambda_1,\lambda_2)$ up to the third order, {\it i.e.} neglecting
a remainder term $r(\lambda_1,\lambda_2)$ which is such that
$|r(\lambda_1,\lambda_2)| \leq C(|\lambda_1|+\lambda_2|)^3$. We
denote such remainders by $\OO(\lambda^3)$ in the sequel. It holds
\begin{eqnarray*}
  \beta \cH(\lambda_1,\lambda_2) - \cS(\lambda_1,\lambda_2) & = &
  \frac{3N}{2} (\lambda_2\delta T - \ln(1+\lambda_2\delta T)) + N \left
  ( \frac{\lambda_1}{2}\left(1 + \frac{1}{1+\lambda_1}\right) - \ln
  (1+\lambda_1) \right ) 
  \\ 
  & & + \lambda_1 \lambda_2 \frac{\delta T}{2}
  \left ( \frac{N}{1+\lambda_1} + \frac{\partial
    A}{\partial_{\lambda_1}}(\lambda_1,\lambda_2) \right ) + \beta
  \lambda_2 \delta T \langle \langle U_{\lambda_1,\lambda_2} \rangle
  \rangle_{\lambda_1,\lambda_2} 
  \\ 
  & & + \frac{\lambda_1}{2} \left
  (\frac{\partial A}{\partial_{\lambda_1}}(0,0) + \frac{\partial
    A}{\partial_{\lambda_1}}(\lambda_1,\lambda_2)\right ) -
  A(\lambda_1,\lambda_2). 
\end{eqnarray*}
Introducing the notation
\[
A_i = \frac{\partial A}{\partial_{\lambda_i}}(0,0), \qquad A_{ij} =
\frac{\partial^2 A}{\partial_{\lambda_i} \partial_{\lambda_j}}(0,0),
\]
the Taylor expansions of the function $A$ and its first derivatives at
an arbitrary state $(\lambda_1,\lambda_2)$ read (using $A(0,0) = 0$):
\[
A(\lambda_1,\lambda_2) = \lambda_1 A_1 + \lambda_2 A_2 +
\frac{\lambda_1^2}{2} A_{11} + \lambda_1 \lambda_2 A_{12} +
\frac{\lambda_2^2}{2} A_{22} + \OO(\lambda^3),
\]
\[
\frac{\partial A}{\partial_{\lambda_i}}(\lambda_1,\lambda_2) = A_i +
\lambda_1 A_{i1} + \lambda_2 A_{i2} + \OO(\lambda^2).
\]
With these Taylor expansions and the relation
\eqref{eq:derivatives_A_2}, it is straightforward to show that
\begin{eqnarray*}
  \beta \cH(\lambda_1,\lambda_2) - \cS(\lambda_1,\lambda_2) & = &
  \frac{3N}{4} \lambda^2_2\delta T^2 + \lambda_1 \lambda_2 \frac{\delta
    T}{2} \left ( N + A_1 \right ) + \lambda_2 (1+\lambda_2\delta T)
  \frac{\partial A}{\partial_{\lambda_2}}(\lambda_1,\lambda_2)\\
  & = & \frac{\lambda_2 \delta T}{2} \left [ \lambda_1 \left ( N + A_1 +
    \frac{A_{12}}{\delta T} \right ) + \lambda_2 \left ( \frac{3N}{2}
    \delta T + 2 A_2 + \frac{A_{22}}{\delta T} \right ) \right ] +
  \OO(\lambda^3).\\
\end{eqnarray*}
Using \eqref{eq:derivatives_A_2}, the derivatives of the entropy
differences can be computed:
\[
\frac{\partial \cS}{\partial \lambda_1}(\lambda_1,\lambda_2) =
\frac{N}{1+\lambda_1} + \frac{\partial A}{\partial
  \lambda_1}(\lambda_1,\lambda_2) + \frac{1+\lambda_2 \delta T}{\delta
  T} \frac{\partial^2 A}{\partial \lambda_1 \partial
  \lambda_2}(\lambda_1,\lambda_2),
\]
\[
\frac{\partial \cS}{\partial \lambda_2}(\lambda_1,\lambda_2) =
\frac{3N \delta T}{2(1+\lambda_2\delta T)} + 2\frac{\partial
  A}{\partial \lambda_2}(\lambda_1,\lambda_2) + \frac{1+\lambda_2
  \delta T}{\delta T} \frac{\partial^2 A}{\partial^2
  \lambda_2}(\lambda_1,\lambda_2).
\]
This shows that
\begin{equation}
\beta \cH(\lambda_1,\lambda_2) - \cS(\lambda_1,\lambda_2) =
\frac{\lambda_2 \delta T}{2} \left ( \lambda_1 \frac{\partial
  \cS}{\partial \lambda_1}(0,0) + \lambda_2 \frac{\partial
  \cS}{\partial \lambda_2}(0,0) \right ) + \OO(\lambda^3),
\end{equation}
so that, since
\[
\cS(\lambda_1,\lambda_2) = \cS(0,0) + \lambda_1 \frac{\partial
  \cS}{\partial \lambda_1}(0,0) + \lambda_2 \frac{\partial
  \cS}{\partial \lambda_2}(0,0) + \OO(\lambda^2),
\]
and $\cS(0,0) = 0$, it holds
\begin{equation}
\beta \cH(\lambda_1,\lambda_2) - \left( 1+\frac{\lambda_2 \delta
  T}{2}\right) \cS(\lambda_1,\lambda_2) = \OO(\lambda^3).
\end{equation}
This relation shows immediately that $\cH(\lambda_1,\lambda_2) =
\OO(\lambda^3)$ on the isentrope, and so, the initial slopes of the
curves, and their first derivatives, coincide.

\section{Precisions on the isentropic integration}

Several numerical schemes may be used to integrate
\eqref{eq:entropy_integration}. The simplest one consists in
approximating the integral appearing in the exponential factor with a
Riemman formula using the value of the integrated function on the left
side of the interval:
\begin{equation}
  \label{eq:T2_estimateur} 
  T_2 \simeq T_1 \exp \left ( -\frac{\partial \cP}{\partial \cU} \bigg
  |_{V_1}(V_2-V_1) \right).
\end{equation}
Of course, higher order integration methods could be used.

It remains to decide how to compute the derivative $\dps
\frac{\partial \cP}{\partial \cU} \bigg |_{V_1}$. Finite differences
may be used to this end, but this would require at least two very
carefully converged simulations with volumes $V_1 \pm \Delta V$.  It
seems more appealing to compute the partial derivative using standard
fluctuations formulas:\cite{lagache01,Bourasseau07}
\begin{equation}
  \label{eq:dEdT}
  \frac{\partial \cU}{\partial T} \bigg |_{V_1} = C_v(V_1,T_1) =
  \frac32 N k_{\rm B} + \frac{1}{k_{\rm B} T_12} \left( \left \langle
  U^2 \right \rangle_{V_1,T_1} - \left \langle U \right
  \rangle^2_{V_1,T_1} \right),
\end{equation}
and
\begin{equation}
  \label{eq:dPdT}
  \frac{\partial \cP}{\partial T} \bigg |_{V_1} = \frac{Nk_{\rm
      B}}{V_1}+\frac{1}{k_{\rm B}T_1^2}\left( \left \langle P H \right
  \rangle_{V_1,T_1} - \left \langle P \right \rangle_{V_1,T_1} \left
  \langle H \right \rangle_{V_1,T_1} \right),
\end{equation}
where $C_v(V_1,T_1)$ is the specific heat at constant volume, and the
pressure observable for a simulation domain of volume $V_1$ reads
\begin{equation}
  \label{eq:pressure}
  P(q,p) = \frac{1}{3 V_1} \sum_{i=1}^N \frac{p_i^2}{m_i} - q_i \cdot
  \nabla_{q_i} U(q).
\end{equation}
The partial derivative $\partial \cP/\partial \cU$ can then be
evaluated in a single simulation at $(N,V_1,T_1)$ using
\eqref{eq:dEdT}-\eqref{eq:dPdT}.

The numerical implementation of this method is done as follows.
The partial derivative of the pressure with respect to the energy 
is first computed with a
Monte Carlo simulation for the given initial conditions $(N,V_1,T_1)$. 
The temperature $T_2$ is then
evaluated from Eq.~\eqref{eq:T2_estimateur}. The partial derivative is
next computed at volume $V_2$ to predict the next
temperature. Proceeding incrementally, the whole isentrope curve
can be constructed. 

The numerical results presented in this work have been obtained 
by performing canonical samplings with a Metropolis algorithm, using the 
Monte-Carlo Gibbs code.\footnote{The Monte-Carlo Gibbs code
is owned by the Institut Francais du P\'etrole, the
Universit\'e Paris-Sud, and the CNRS, and developped in collaboration
with the CEA.
} Partial derivatives have been computed in
the NVT ensemble. The convergence of simple thermodynamic averages
was generally obtained after $N_{\rm steps} = 10^7$ iterations, but
derivative properties (related to the covariance of some observables) required 
about $N_{\rm steps} = 10^8$ iterations for a satisfactory convergence. 
Error estimates on the canonical
samplings have been obtained with block averaging,\cite{FP89} and the
error propagation estimated along the integration scheme has been
computed using standard propagation rules. In all the cases
considered, the statistical error (as measured using the 95\%
confidence interval associated with the variance computed from
block-averaging) on the predicted temperature on the isentrope curve
is inferior to 1.5 \%.


\bibliography{biblio}


\newpage

Caption list: \\

\begin{figure}
\centering
 \begin{minipage}{15.0cm}
    \begin{tabular}{cc}
      \includegraphics*[width=7.0cm]{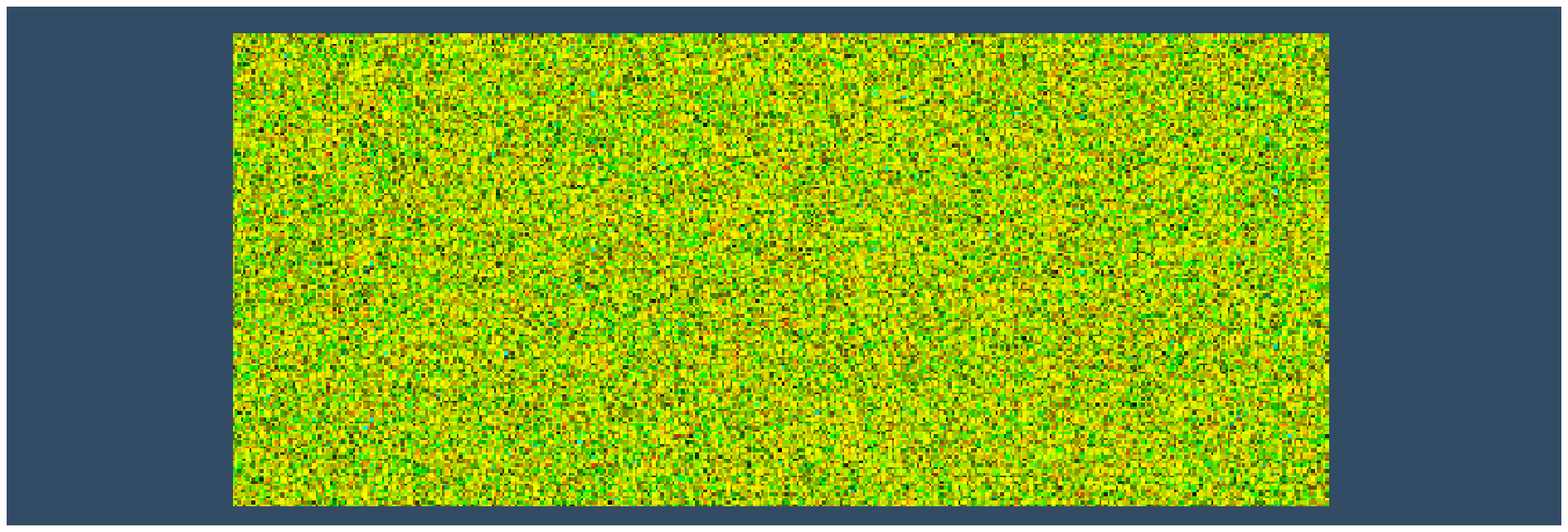}&
      \includegraphics*[width=7.0cm]{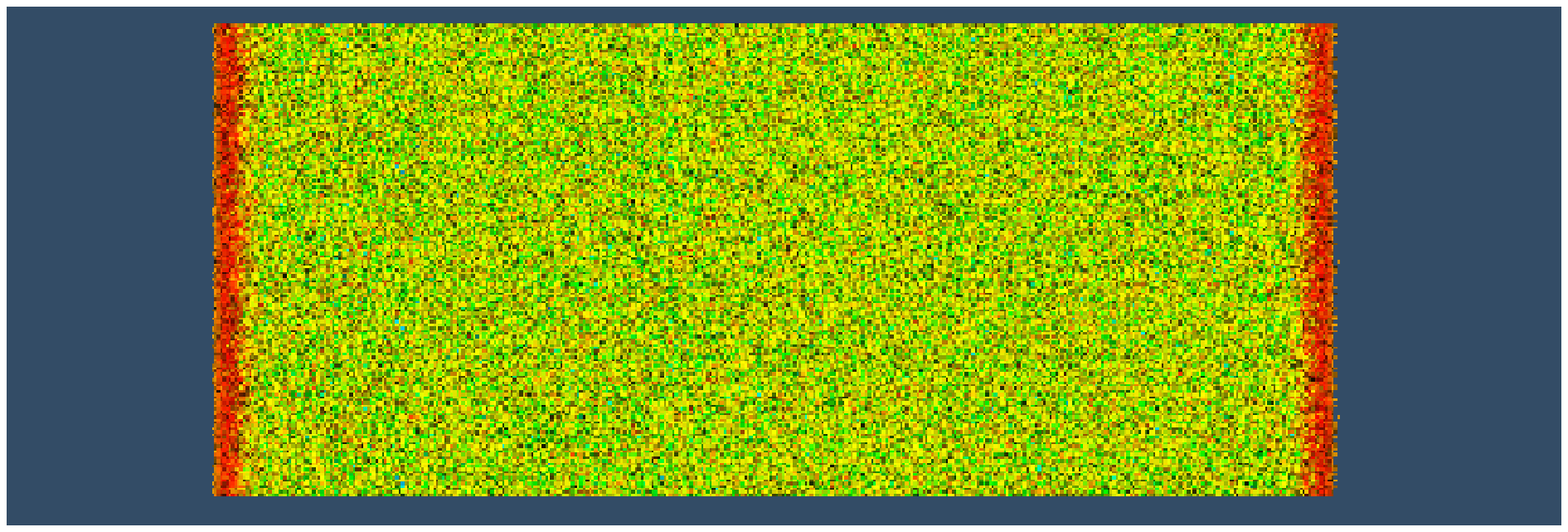}\\ 
      \includegraphics*[width=7.0cm]{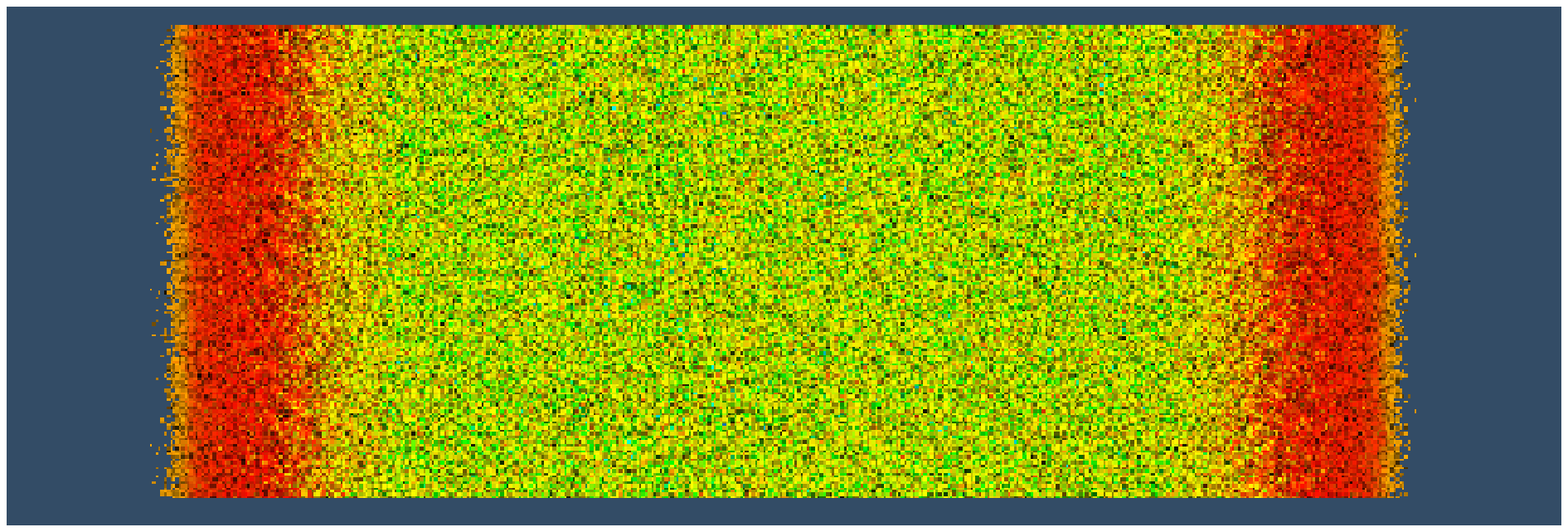}&
      \includegraphics*[width=7.0cm]{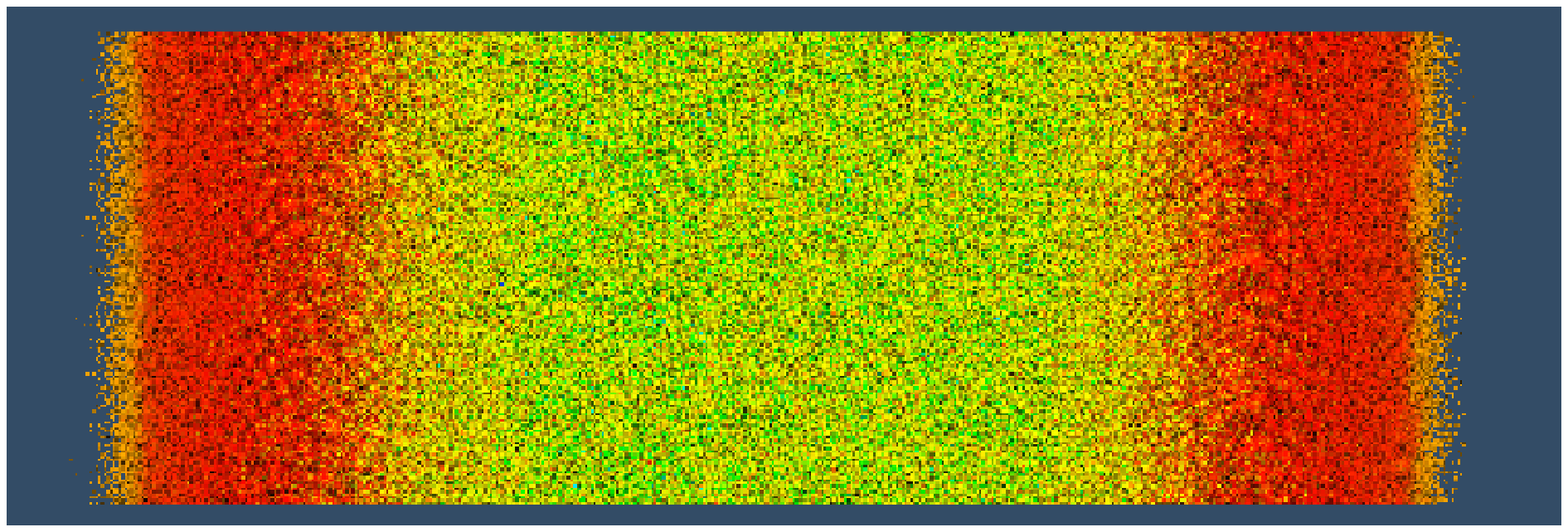}\\
    \end{tabular}
  \end{minipage}
\caption{}\label{NEMDprocess} (a)
\end{figure}

Figure \ref{NEMDprocess}: (a) (color online) Non-Equilibrium Molecular Dynamics of
isentropic release waves. The four pictures represent snapshots of the
system during the release process, the expansion
proceeding in the longitudinal $x$ direction. Atoms are colored according to
their potential energies (scaling corresponding to $-1.38 \times 10^{-20}$~J 
for blue up to $7.55 \times 10^{-20}$~J for red).\\

Figure \ref{NEMDprocess}: (b) (color online) Density profiles taken at different
times of the simulation (from blue to red as time increases).\\

\begin{figure}
\centering
\includegraphics[width=14cm]{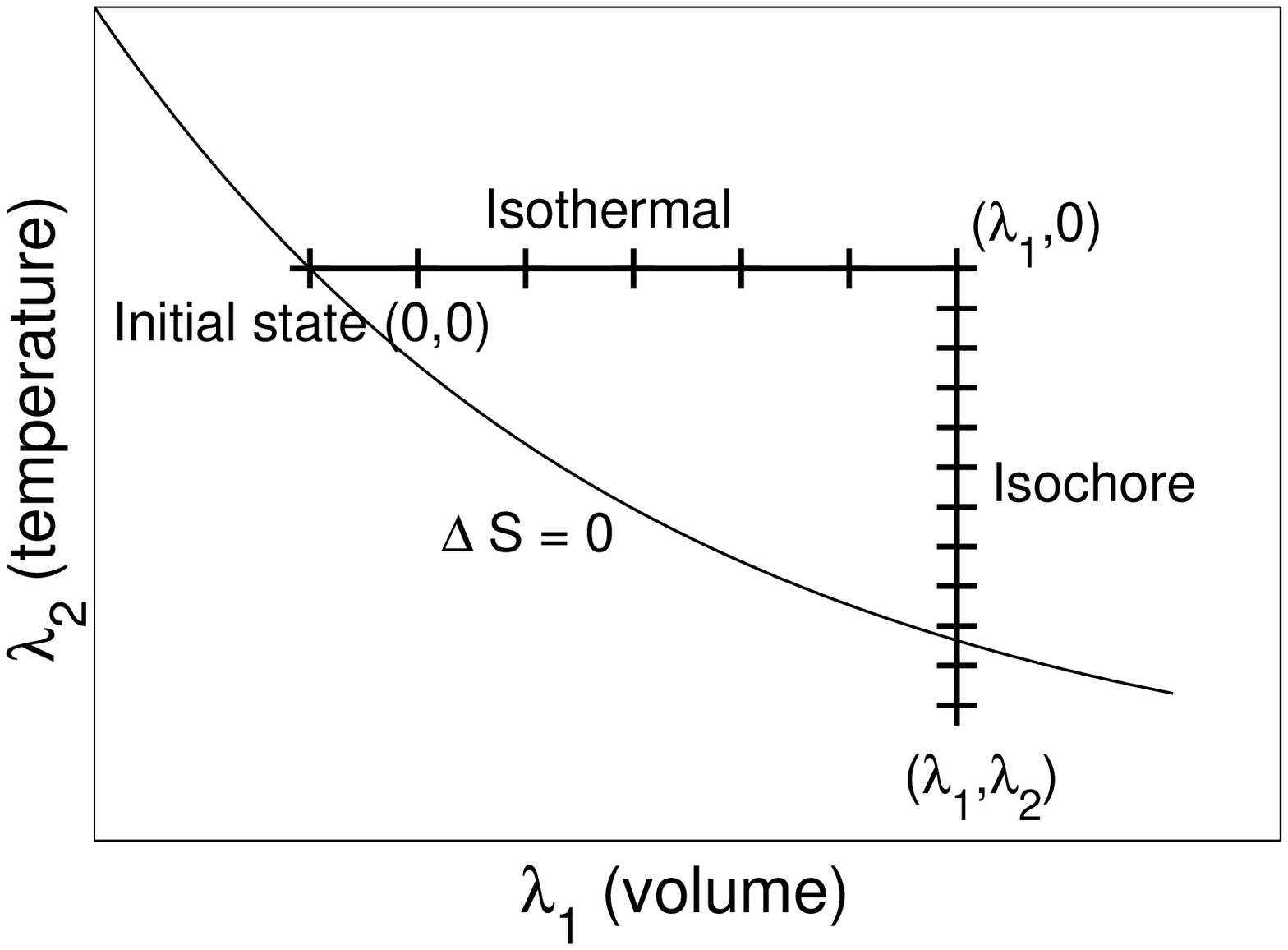}
\caption{}\label{fig:TI_path}
\end{figure}

Figure \ref{fig:TI_path}: Path in the $(\lambda_1,\lambda_2)$ space used to compute
states with the same entropy as the initial state.
Each cross represents some equilibrium canonical sampling along the thermodynamic path.  
First, the isothermal expansion is performed (horizontal line in the diagram), 
starting from the initial state $(0,0)$, 
until the required density is reached.
The entropy of the state $(\lambda_1,0)$ is $S_{\rm init} + \Delta S_{\rm expansion}$. 
Then, an isochore cooling is performed ($\lambda_1$ is kept fixed; vertical line in the
diagram), until the entropy difference
during this process is the opposite of the entropy variation found in the expansion part.
The final state $(\lambda_1,\lambda_2)$, located
at the intersection of the curve $\Delta S = 0$ and the vertical line,
has then the same entropy as the initial state.\\
  
\begin{figure}
\centering
\epsfig{file=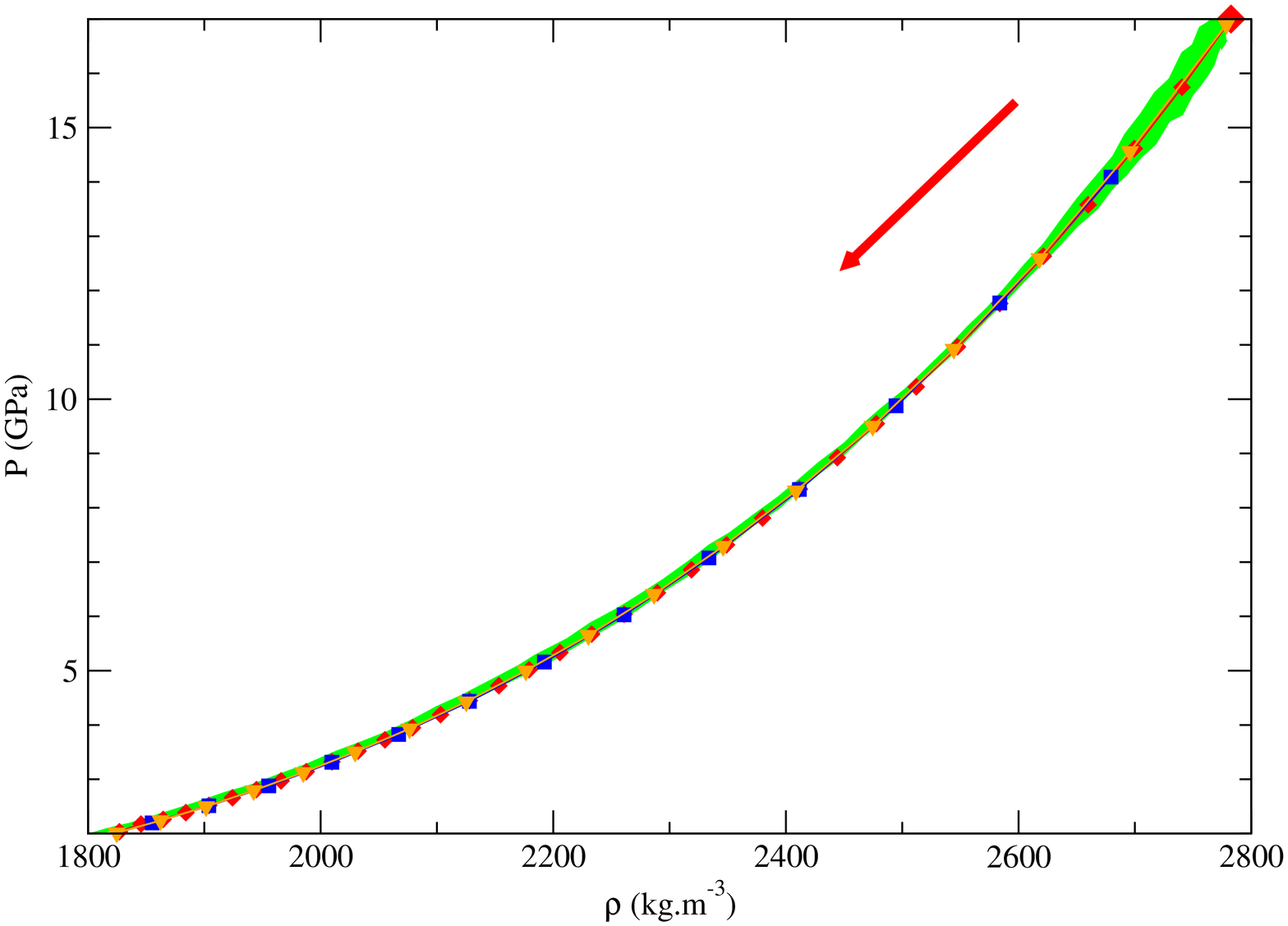,width=14cm,angle=270}
\caption{}\label{P_rho}
\end{figure}

Figure \ref{P_rho}: (color online) Isentropic release in a ($P, \rho$)
diagram. Symbols represent results from equilibrium methods, red
diamonds for the successive hugoniostat, blue squares for the
thermodynamic integration and yellow triangles for the entropy
integration. NEMD results are plotted in green, the width of the 
so-obtained tube corresponding to the error bars. 
The arrow indicates the path followed during the release.\\

\begin{figure}
\centering
\epsfig{file=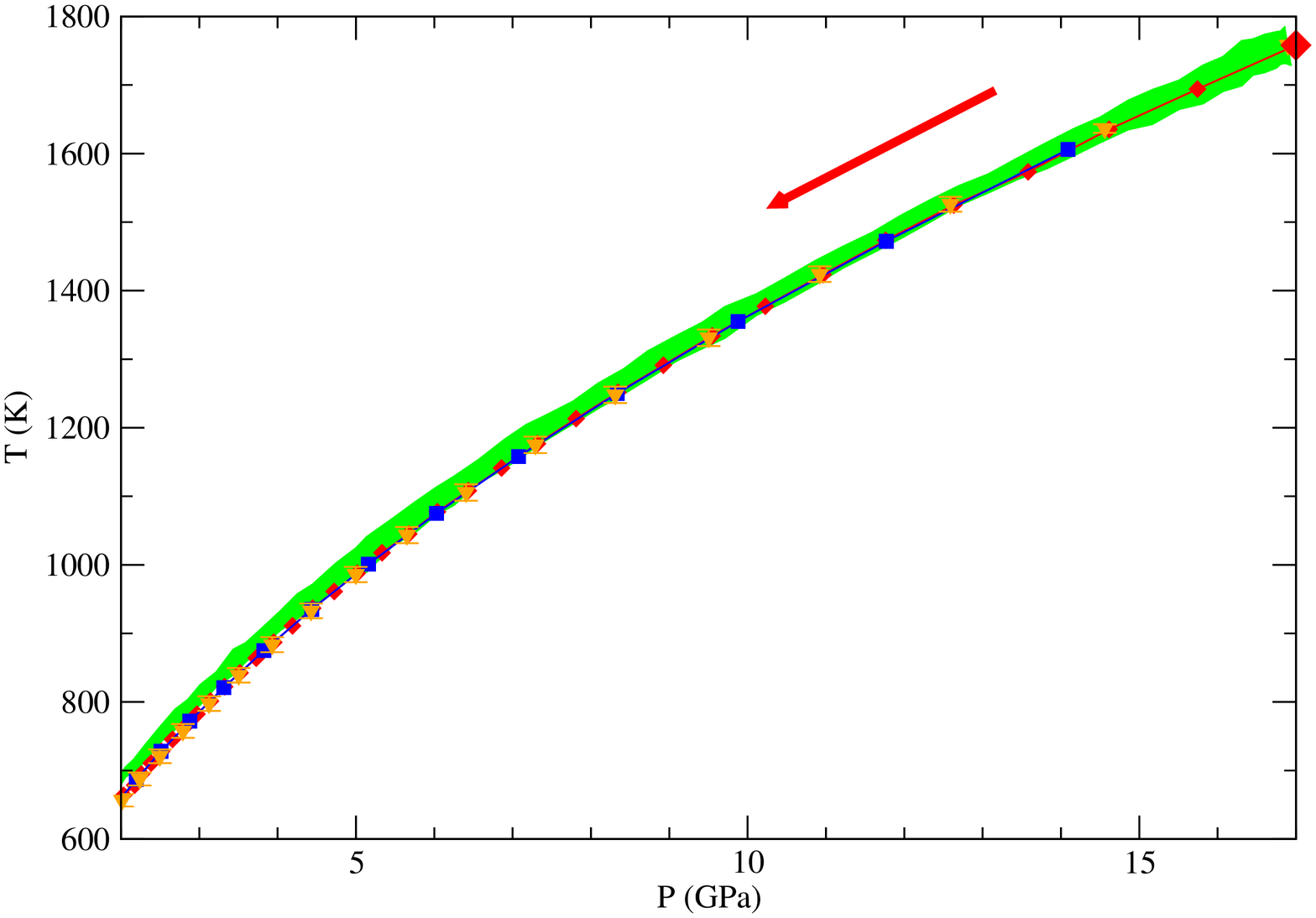,width=14cm,angle=270}
\caption{}\label{T_P}
\end{figure}

Figure \ref{T_P}: (color online) Isentropic release in a ($T, P$) diagram. 
The symbols are the same as in Figure \ref{P_rho}. 
Notice that there is slight deviation of the NEMD results for the lowest temperatures.\\

\begin{figure}
\centering
\epsfig{file=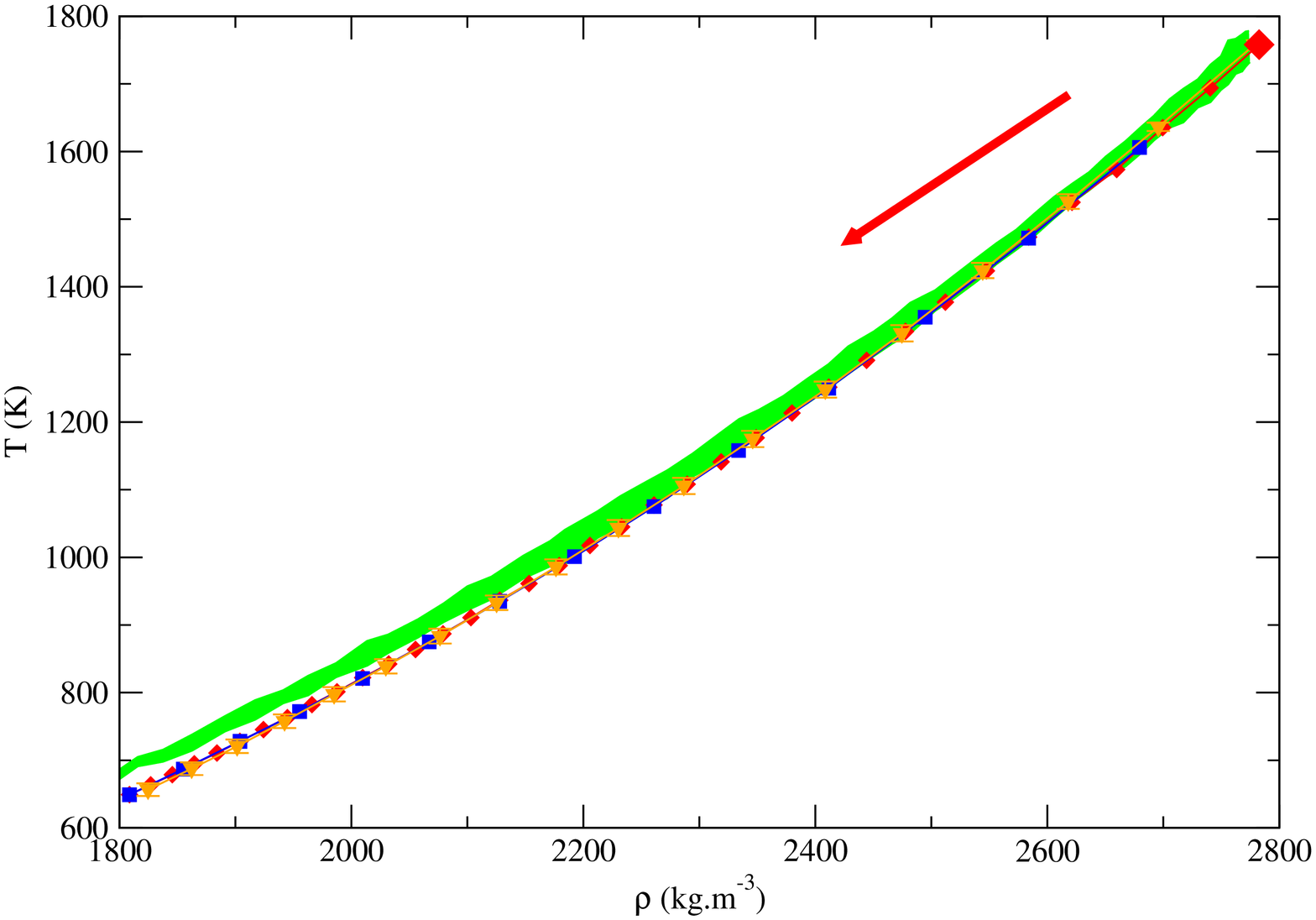,width=14cm,angle=270}
\caption{}\label{T_rho}
\end{figure}

Figure \ref{T_rho}: (color online) Isentropic release in a ($T, \rho$)
diagram. The symbols are the same as in Figure \ref{P_rho}.
There is a noticeable deviation of the NEMD results for the lowest temperatures.
\\

\newpage

\newpage

\begin{figure}[h!]
\centering
\epsfig{file=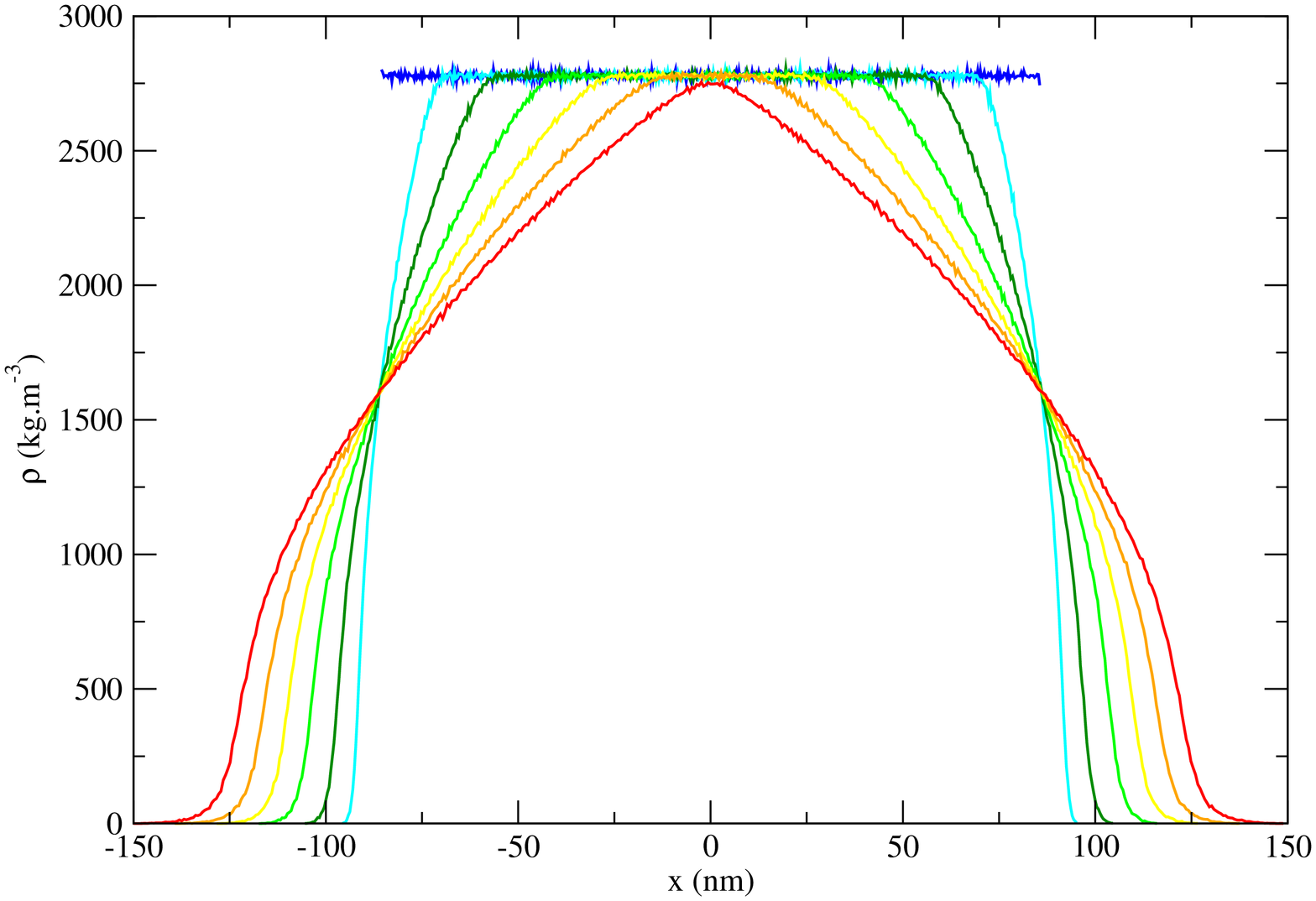,width=14cm,angle=270}\\
Fig~\ref{NEMDprocess}: (b)
\end{figure}

\newpage

\newpage

\newpage

\newpage

\end{document}